\documentclass[12pt,english,preprint,showpacs,preprintnumbers,nofootinbib]{revtex4}
\usepackage[T1]{fontenc}
\usepackage[latin1]{inputenc}
\usepackage{babel}
\usepackage{color}
\usepackage{graphics}

\makeatletter

\providecommand{\LyX}{L\kern-.1667em\lower.25em\hbox{Y}\kern-.125emX\@}
\let\SF@@footnote\footnote
\def\footnote{\ifx\protect\@typeset@protect
    \expandafter\SF@@footnote
  \else
    \expandafter\SF@gobble@opt
  \fi
}
\expandafter\def\csname SF@gobble@opt \endcsname{\@ifnextchar[%]
  \SF@gobble@twobracket
  \@gobble
}
\edef\SF@gobble@opt{\noexpand\protect
  \expandafter\noexpand\csname SF@gobble@opt \endcsname}
\def\SF@gobble@twobracket[#1]#2{}

\usepackage{psfig}
\usepackage{graphicx}% Include figure files
\usepackage{dcolumn}% Align table columns on decimal point
\usepackage{bm}% bold math
\usepackage{amssymb}
\usepackage{here}
\usepackage{mathrsfs}

\def\ra{\rightarrow}

\def\be{\begin{equation}}
\def\ee{\end{equation}}
\def\bea{\begin{eqnarray}}
\def\eea{\end{eqnarray}}
\def\bea{\begin{eqnarray}}
\def\ena{\end{eqnarray}}

\def\ra{\rightarrow}

\def\bea{\begin{eqnarray*}}
\def\ena{\end{eqnarray*}}
\def\beq{\begin{equation*}}
\def\enq{\end{equation*}}

\def\D0{D\O~}

\def\today{\number\day
    \space\ifcase\month\or
      January\or February\or March\or April\or May\or June\or
      July\or August\or September\or October\or November\or
             December\fi
    \space\number\year}

\makeatother
\begin{document}
\thispagestyle{empty} \setcounter{page}{0}
\setcounter{footnote}{1} \renewcommand{\thefootnote}{\fnsymbol{footnote}}

\preprint{hep-ph/0212159}
\title{Tevatron Run-1 $Z$ Boson Data and \\
 Collins-Soper-Sterman Resummation Formalism}

\author{F. Landry$ ^{1} $, R. Brock$ ^{1} $, P.~M. Nadolsky$ ^{2} $ and
C.--P. Yuan$ ^{1} $}

\email{brock@pa.msu.edu; nadolsky@mail.physics.smu.edu; yuan@pa.msu.edu}

\affiliation{\vspace*{2mm} $ ^{1} $ Department of Physics and Astronomy, Michigan
State University, E. Lansing, MI 48824, USA \\
 $ ^{2} $ Department of Physics, Southern Methodist University, Dallas,
TX 75275-0175, USA}

\date{\today{}}

\begin{abstract}
We examine the effect of the $ Z $-boson transverse momentum distribution
measured at the Run-1 of the Tevatron on the nonperturbative function of
the Collins-Soper-Sterman (CSS) formalism, which resums large logarithmic
terms from multiple soft gluon emission in hadron collisions. The inclusion
of the Tevatron Run-1 $ Z $ boson data strongly favors a Gaussian form
of the CSS nonperturbative function, when combined with the other low energy
Drell-Yan data in a global fit.
\end{abstract}


\maketitle
\newpage
\setcounter{page}{1} \pagenumbering{arabic} \pagestyle{plain}

\setcounter{footnote}{0} \renewcommand{\thefootnote}{\arabic{footnote}}

\section{Introduction}

In hadron-hadron collisions, the transverse momentum of
Drell-Yan pairs or weak gauge bosons ($ W^{\pm } $ and $ Z $) is generated
by emission of gluons and quarks, as predicted by Quantum Chromodynamics
(QCD). Therefore, in order to test the QCD theory or electroweak properties
of the vector bosons, it is necessary to include the effects of multiple
gluon emission. One theoretical framework designed to account for these effects
is the resummation formalism developed by Collins, Soper, and Sterman (CSS)
\cite{CSS}, which has been applied to study production of
single \cite[and references therein]{AK, oneW, wres, ERV}
and double \cite{twoW1,twoW2} electroweak gauge bosons, as well as Higgs
bosons \cite{Higgs}, at hadron colliders. Just as the nonperturbative functions,
\emph{i.e.}, parton distribution functions
(PDF's), are needed in order to predict inclusive rates, an additional nonperturbative
function, $ \widetilde{W}_{j\bar{k}}^{NP} $, is required in the CSS formalism
to describe the transverse momentum of, say, weak bosons. Many studies have
been performed in the literature to determine $ \widetilde{W}_{j\bar{k}}^{NP} $
using the available low energy Drell-Yan data \cite{DWS,LY,ERV,BLLY,qiu1,qiu2}.
In particular, in Ref. \cite{BLLY} three of us have examined various functional
forms of $ \widetilde{W}_{j\bar{k}}^{NP} $ to test the universality of
the CSS formalism in describing the Drell-Yan and weak boson data. The result
of that study was summarized in Table II of Ref. \cite{BLLY}. In addition,
that paper made several important observations, as to be discussed below.
First, neither the DWS form (cf.~Eq.~(\ref{DWS_form})) nor the LY form
(cf.~Eq.~(\ref{LY_form})) of the nonperturbative
function $ \widetilde{W}_{j\bar{k}}^{NP} $
could simultaneously describe the Drell-Yan
data in a straightforward global fit of
 the experiments R209 \cite{R209}, E605 \cite{E605}, and E288
\cite{E288}, as well as the Tevatron Run 0 $ Z $ boson data from the
CDF collaboration \cite{CDFZ}. Hence, it was decided in Ref. \cite{BLLY}
to first fit only the first two mass bins ($ 7<Q<8 $\,GeV and $ 8<Q<9 $\,GeV)
of the E605 data and all of the R209 and the CDF-$ Z $ boson data in the
initial fits $ A_{2} $ and $ A_{3} $. In total, 31 data points were
considered. We allowed the normalization of the R209 and E605 data to float
within their overall systematic normalization errors, while fixing the normalization
of the CDF-$ Z $ Run 0 data to unity. Second, after the initial fits,
we calculated the remaining three high-mass bins of the E605 data not used
in Fits $ A_{2,3} $ and found a reasonable agreement with the experimental
data. In order to compare with the E288 data, we created Fits $ N_{2,3} $,
in which we fixed the functions $ \widetilde{W}_{j\bar{k}}^{NP} $ to those
obtained from Fit $ A_{3} $ and performed a fit for NORM (the fitted normalization
factor applied to the prediction curves for a given data set) for the E288
data alone. We found that the quality of the fit for the E288 data is very
similar to that for the E605 data, and that the normalizations were then acceptably
within the range quoted by the experiment. Hence, we concluded that the fitted
functions $ \widetilde{W}_{j\bar{k}}^{NP} $ reasonably describe the wide-ranging,
complete set of data, in the sense discussed above. Third, most importantly,
we found that the complete set of data available in that fit was not yet
precise enough to clearly separate the $ g_{2} $ and $ g_{1}g_{3} $
parameters within a pure Gaussian form of $ \widetilde{W}_{j\bar{k}}^{NP} $
with $ x $ dependence similar to that of the LY form. This Gaussian form
is given explicitly in Eq.~(\ref{BLNY_form}).

\textcolor{black}{In this paper, we show that, after including the transverse
momentum distributions of the $ Z $ bosons measured by the \D0 \cite{run1D0}
and CDF collaborations \cite{run1CDF} in Run-1 at the Fermilab Tevatron,
we are able for the first time to perform a truly global fit of the nonperturbative
function $ \widetilde{W}_{j\bar{k}}^{NP} $ to the complete set of data
on vector boson production. In this fit, the data from the experiments R209,
E605, and E288, as well as the Tevatron Run-1 $ Z $ data are treated on
the same footing. We emphasize that in this new fit, the E288 data are also
included in the global fit, in contrast to the study done in Ref. \cite{BLLY}.
Furthermore, we show that the Gaussian form of $ \widetilde{W}_{j\bar{k}}^{NP} $
given in Eq.~(\ref{BLNY_form}) clearly fits the data the best, as compared
to either the updated DWS form (\ref{DWS_form}) or updated LY form (\ref{LY_form}).
These nice features are driven by the inclusion of the Run~1 $ Z $~data,
for these data determine the $ g_{2} $ coefficient with good accuracy
by separating the contributions from $ g_{2} $ and $ g_{1}g_{3} $. }

\textcolor{black}{This paper is organized as follows. In Section II, we briefly
review the CSS formalism with an emphasis on its nonperturbative sector.
In Section III, we describe in detail the results of our fits. In Section
IV, we discuss various aspects of our study and comment on the validity of
our approach to the treatment of the nonperturbative region. Section V contains
conclusions.}

\section{Collins-Soper-Sterman Resummation Formalism}

\textcolor{black}{As an example, we consider production of a vector boson
$ V $ in the collision of two hadrons $ h_{1} $ and $ h_{2} $. In
the CSS resummation formalism, the cross section for this process is written
in the form\begin{equation}
\label{WY}
{\frac{d\sigma (h_{1}h_{2}\rightarrow VX)}{dQ^{2}\, dQ_{T}^{2}dy\, \, }}=
{\frac{1}{(2\pi )^{2}}}\delta \left( Q^{2}-M_{V}^{2}\right) \int d^{2}b\,
e^{i{\vec{Q}_{T}}\cdot {\vec{b}}}{\widetilde{W}_{j\bar{k}}(b,Q,x_{1},x_{2})}+
Y(Q_{T},Q,x_{1},x_{2}),
\end{equation}
 where $ Q,\, Q_{T}, $ and $ y $ are the invariant mass, transverse
momentum, and the rapidity of the vector boson $ V $. The Born-level parton
momentum fractions are defined as $ x_{1}=e^{y}Q/\sqrt{S} $ and $ x_{2}=e^{-y}Q/\sqrt{S} $,
with $ \sqrt{S} $ being the center-of-mass (CM) energy of the hadrons
$ h_{1} $ and $ h_{2} $. We will refer to the integral over the impact
parameter $ b $ in Eq.~(\ref{WY}) as the {}``$ \widetilde{W} $-term{}''.
$ Y $~is the regular piece, which can be obtained by subtracting the
singular terms from the exact fixed-order result. The quantity $ \widetilde{W}_{j\bar{k}} $
satisfies a set of renormalization- and gauge-group equations \cite{CS}
with the solution of the form\begin{equation}
\widetilde{W}_{j\bar{k}}(b,Q,x_{1},x_{2})=e^{-{\mathcal{S}}(Q,b{,C_{1},C_{2}})}
\widetilde{W}_{j\bar{k}}\left( b,\frac{C_{1}}{C_{2}b},x_{1},x_{2}\right) ,
\end{equation}
 where $ C_{1} $ and $ C_{2} $ are constants of order unity, and the
Sudakov exponent is defined as\begin{equation}
{\mathcal{S}}(Q,b{,C_{1},C_{2}})=\int _{C_{1}^{2}/b^{2}}^{C_{2}^{2}Q^{2}}
\frac{d\overline{\mu }^{2}}{\overline{\mu }^{2}}\left[ {\cal A}
\left( {\alpha }_{s}(\overline{\mu }),C_{1}\right)
 \ln \left( \frac{C_{2}^{2}Q^{2}}{\overline{\mu }^{2}}\right) +
 {\cal B}\left( {\alpha }_{s}(\overline{\mu }),C_{1},C_{2}\right) \right] .
\end{equation}
 The dependence of $ \widetilde{W}_{j\bar{k}} \left( b,\frac{C_{1}}{C_{2}b},x_{1},x_{2}\right) $
on $ x_{1} $ and $ x_{2} $
factorizes as \begin{equation}
\widetilde{W}_{j\bar{k}}\left( b,\frac{C_{1}}{C_{2}b},x_{1},x_{2}\right) =
\sum _{j,\bar{k}}\frac{\sigma _{0}}{S}\; {\overline{{\mathscr P}}}_{jh_{1}}
\left( x_{1},b,\frac{C_{1}}{C_{2}b}\right) \; {\overline{{\mathscr P}}}_{\bar{k}h_{2}}
\left( x_{2},b,\frac{C_{1}}{C_{2}b}\right) +(j\leftrightarrow \bar{k}).
\end{equation}
 In the perturbative region,} \textcolor{black}{\emph{i.e.}}\textcolor{black}{,
at $ b^{2}\ll 1/\Lambda ^{2}_{QCD}, $ the function $
{\overline{{\mathscr P}}}_{jh} $ can be expressed as a
convolution of the parton distribution functions $ f_{a/h} $ with
calculable Wilson coefficient functions $ {\cal C}_{ja} $:
\begin{equation} \label{mathscrP} {\overline{{\mathscr
P}}}_{jh}\left( x,b,\frac{C_{1}}{C_{2}b}\right) = \sum _{a}\int
_{x}^{1}{\frac{d\xi }{\xi }}\; {\cal C}_{ja}\left( \frac{x}{\xi
},b, \frac{C_{1}}{C_{2}b},\mu =\frac{C_{3}}{b}\right) \;
f_{a/h}\left( \xi ,\mu=\frac{C_{3}}{b}\right) .
\end{equation}
 The sum over the index $ a $ is over all types of incoming partons. The
sum over the index $ j $($ \bar{k} $) is over all quarks (antiquarks).
The coefficient $ \sigma _{0} $ includes constant factors and quark couplings
from the leading-order cross section, which can be found,}
\textcolor{black}{\emph{e.g.}}\textcolor{black}{,
in Ref.~\cite{CSS}. The factorization scale $ \mu  $ on the r.h.s. of
Eq.~(\ref{mathscrP}) is fixed to be $ C_{3}/b $. }

\textcolor{black}{A few comments should be made about this formalism: }

\begin{itemize}
\item \textcolor{black}{If both $ Q $ and $ 1/b $ are much larger than the
typical internal hadronic scale $ \Lambda _{QCD} $, the $ {\cal A} $,
$ {\cal B} $ and $ {\cal C} $ functions can be calculated order-by-order
in $ \alpha _{s} $. In our fit, we shall include the $ {\cal A} $ and
$ {\cal B} $ functions up to $ {\mathcal{O}}(\alpha _{s}^{2}) $, and
$ {\cal C} $ functions up to $ {\mathcal{O}}(\alpha _{s}) $. }
\item \textcolor{black}{A special choice can be made for the renormalization constants
$ C_{i} $ to remove some of the logarithms in $ \widetilde{W}_{j\bar{k}}(b,Q,x_{1},x_{2}) $.
This canonical choice is $ C_{1}=C_{3}=2e^{-\gamma _{E}}\equiv b_{0} $
and $ C_{2}=C_{1}/b_{0}=1 $, where $ \gamma _{E} $ is the Euler's constant.
We shall use this choice in our calculations. }
\item \textcolor{black}{In Eq.~(\ref{WY}), the variable $ b $ is integrated
from 0 to $ \infty  $. When $ b\gtrsim 1\mbox {\, GeV}^{-1} $, the
perturbative calculation for $ \widetilde{W}_{j\bar{k}}(b,Q,x_{1},x_{2}) $
is no longer reliable, and complicated long-distance physics comes in. Furthermore,
even in the perturbative region ($ b\lesssim 1\mbox {\, GeV}^{-1} $)
$ \widetilde{W}_{j\bar{k}}(b,Q,x_{1},x_{2}) $
may contain some small nonperturbative terms, which arise,}
\textcolor{black}{\emph{e.g.}}\textcolor{black}{,
from power corrections to the CSS evolution equations. It is important to
emphasize that significance of nonperturbative contributions of both types
is drastically reduced when $ Q $ is of order of the $ W,\, Z $ boson
masses or higher \cite{CSS}. For those $ Q $, the most part of the $ Q_{T} $
distribution can be predicted based purely on the perturbative calculation,
with the exception of the region of $ Q_{T} $ below a few GeV, where sizeable
sensitivity to the nonperturbative input remains \cite{DWS,AK}. }
\end{itemize}
\textcolor{black}{According to the common assumption, nonperturbative contributions
to $ \widetilde{W}_{j\bar{k}}(b,Q,x_{1},x_{2}) $ can be approximated by
some phenomenological model with} \textcolor{black}{\emph{}}\textcolor{black}{measurable
and universal}\footnote{%
Here, we mean {}``universal{}'' in the context of
\textcolor{black}{Drell-Yan-like processes},
in which the initial state of the Born level process involves
only quarks and antiquarks, and the \textcolor{black}{observed final state
does not participate in strong interactions. }
} \textcolor{black}{parameters. Collins, Soper, and Sterman \cite{CSS} suggested
the introduction of the nonperturbative terms in the form of an additional factor
$ \widetilde{W}^{NP}_{j\bar{k}}(b,Q,x_{1},x_{2}) $, usually referred to
as the {}``nonperturbative Sudakov function{}''. More precisely, the form
factor $ \widetilde{W}_{j\bar{k}}(b,Q,x_{1},x_{2}) $ in Eq.~(\ref{WY})
is expressed in terms of its perturbative part $ \widetilde{W}^{pert}_{j\bar{k}} $
and nonperturbative function $ \widetilde{W}^{NP}_{j\bar{k}} $
as\footnote{Here and after, we suppress the arguments $Q$, $x_{1}$ and
$x_{2}$, and denote 
$ \widetilde{W}_{j\bar{k}}(b,Q,x_{1},x_{2}) $
as
$ \widetilde{W}_{j\bar{k}}(b)$, etc.
}
\begin{equation}
\label{wpres}
\widetilde{W}_{j{\bar{k}}}(b)=\widetilde{W}_{j{\bar{k}}}^{pert}(b_{*})
\widetilde{W}_{j{\bar{k}}}^{NP}(b)\, ,
\end{equation}
 with\begin{equation}
\label{bstar}
b_{*}={\frac{b}{\sqrt{1+(b/b_{max})^{2}}}}.
\end{equation}
 In numerical calculations, $ b_{max} $ is typically set to be of order
of $ 1\mbox {\, GeV}^{-1} $. The variable $ b_{*} $ never exceeds $ b_{max} $,
so that $ \widetilde{W}^{pert}_{j{\bar{k}}}(b_{*}) $ can be reliably calculated
in perturbation theory for all values of $ b $. Based upon the renormalization
group analysis, Ref.~\cite{CSS} found that the nonperturbative function
can be generally written as \begin{equation}
\widetilde{W}_{j\bar{k}}^{NP}(b,Q,Q_{0},x_{1},x_{2})=
\exp \left[ -F_{1}(b)\ln \left( \frac{Q^{2}}{Q_{0}^{2}}\right) -
F_{j/{h_{1}}}(x_{1},b)-F_{{\bar{k}}/{h_{2}}}(x_{2},b)\right] \, ,
\end{equation}
 where $ F_{1} $, $ F_{j/{h_{1}}} $ and $ F_{{\bar{k}}/{h_{2}}} $
must be extracted from the data, with the constraint that \begin{equation}
\widetilde{W}_{j\bar{k}}^{NP}(b=0)=1.
\end{equation}
 Furthermore, $ F_{1} $ depends only on $ b $. $ F_{j/{h_{1}}} $
and $ F_{{\bar{k}}/{h_{2}}} $ in general depend on $ x_{1} $ or $ x_{2} $,
and their values can depend on the flavor of the initial-state partons ($ j $
and $ \bar{k} $ in this case). Later, it was shown in Ref. \cite{sterman}
that the $ F_{1}(b)\ln \left( \frac{Q^{2}}{Q_{0}^{2}}\right)  $ dependence
is also suggested by infrared renormalon contributions to the form factor
$ \widetilde{W}_{j\bar{k}}(b,Q,x_{1},x_{2}) $. The CSS resummation formalism
suggests that the nonperturbative function is universal. Its role is analogous
to that of the parton distribution function in any fixed-order perturbative
calculation. In particular, its origin is due to the long-distance effects
that are incalculable at the present time, and its value must be determined
from data. }

\textcolor{black}{As discussed in Ref. \cite{BLLY}, we will consider three
different functional forms for $ \widetilde{W}_{j\bar{k}}^{NP} $ . They
are }

\begin{itemize}
\item \textcolor{black}{the 2-parameter pure Gaussian form,
called the Davies-Webber-Stirling
(DWS) form \cite{DWS}, \begin{equation}
\label{DWS_form}
{\textrm{exp}}\left[ -g_{1}-g_{2}\ln \left( {\frac{Q}{2Q_{0}}}\right) \right] b^{2};
\end{equation}
 }
\item \textcolor{black}{the Ladinsky-Yuan (LY) form \cite{LY}, \begin{equation}
\label{LY_form}
{\textrm{exp}}\left\{ \left[ -g_{1}-g_{2}\ln \left( {\frac{Q}{2Q_{0}}}\right)
\right] b^{2}-\left[ g_{1}g_{3}\ln \! {(100x_{1}x_{2})}\right] b\right\} ,
\end{equation}
 which has a logarithmic $ x $-dependent term linear in $ b $; }
\item \textcolor{black}{and the 3-parameter pure Gaussian form, called the
Brock-Landry-Nadolsky-Yuan (BLNY) form,
\begin{equation}
\label{BLNY_form}
{\textrm{exp}}\left[ -g_{1}-g_{2}\ln \left( {\frac{Q}{2Q_{0}}}\right) -
g_{1}g_{3}\ln {(100x_{1}x_{2})}\right] b^{2}.
\end{equation}
}
\end{itemize}
We will refer to the updated DWS and LY parameterizations obtained in the
current \emph{}\textit{\emph{global}} fit as {}``DWS-G{}'' and {}``LY-G{}''
parameterizations, respectively, to distinguish them from the original DWS
and LY parameterizations \cite{DWS,LY} obtained in (non-global)
fits to a part of the current data.

\section{\label{sec:fits}Results of the Global fits}

\textcolor{black}{In order to examine the impact of including the $ Z $
boson data from the Run-1 of the Tevatron on the global fits and compare
the new results to those given in Ref.~\cite{BLLY}, our theory calculations
will consistently use CTEQ3M PDF's \cite{cteq3}.}\footnote{%
In principle, the non-perturbative function depends on the choice of the
PDF's. \textcolor{black}{However, we will argue later that this dependence
can be currently neglected within the accuracy of the existing data.}
} \textcolor{black}{For the same reason, we take $ Q_{0}=1.6\mbox {\, GeV} $
and $ b_{max}=0.5\mbox {\, GeV}^{-1} $ in all fits.}

\begin{table}

\caption{\textcolor{black}{Vector boson production} data used in this analysis. Here,
\protect$ \delta N_{exp}\protect $ is the published normalization uncertainty
for each experiment. }

{\centering \begin{tabular}{|c|c|c|c|c|}
\hline
Experiment &
 Reference &
 Reaction &
 $ \sqrt{S} $ (GeV) &
 $ \delta N_{exp} $\\
\hline
\hline
R209 &
 \cite{R209}&
 $ p+p\ra \mu ^{+}\mu ^{-}+X $&
 62 &
 10\%\\
\hline
E605 &
 \cite{E605}&
 $ p+Cu\ra \mu ^{+}\mu ^{-}+X $&
 38.8 &
 15\%\\
\hline
E288 &
 \cite{E288}&
 $ p+Cu\ra \mu ^{+}\mu ^{-}+X $&
 27.4 &
 25\%\\
\hline
CDF-$ Z $&
 \cite{CDFZ}&
 $ p+\bar{p}\ra Z+X $&
 1800 &
 -- \\
 (Run-0) &
&
&
&
\\
\hline
\D0-$ Z $&
 \cite{run1D0}&
 $ p+\bar{p}\ra Z+X $&
 1800 &
 4.3\%\\
 (Run-1) &
&
&
&
\\
\hline
CDF-$ Z $&
 \cite{run1CDF}&
 $ p+\bar{p}\ra Z+X $&
 1800 &
 3.9\% \\
 (Run-1) &
&
&
&
\\
\hline
\end{tabular}\par}

{\centering \label{data_table}\par}
\end{table}

\begin{table}

\caption{The data sets used for the fit. \protect$ {\textrm{P}}_{T}\protect $
and \protect$ Q\protect $ denote the \textcolor{black}{published} transverse
momentum and mass of the Drell-Yan pair or the \protect$ Z\protect $ boson,
respectively.}

{\centering \begin{tabular}{|c|c|c|}
\hline
Experiment &
 $ {\textrm{P}}_{T} $ range &
 $ Q $ range \\
&
 (GeV) &
 (GeV) \\
\hline
\hline
R209 &
 0.0 - 1.8 &
 5.0 - 11.0 \\
\hline
E605 &
 0.0 - 1.4 &
 7.0 - 9.0 \& 10.5 - 18.0 \\
\hline
E288 &
 0.0 - 1.4 &
 5.0 - 9.0 \\
\hline
CDF-$ Z $&
 0.0 - 22.8 &
 91.19 \\
 (Run-0) &
&
\\
\hline
\D0-$ Z $&
 0.0 - 22.0 &
 91.19 \\
 (Run-1) &
&
\\
\hline
CDF-$ Z $&
 0.0 - 22.0 &
 91.19 \\
 (Run-1) &
&
 \\
\hline
\end{tabular}\par}

{\centering \label{data_bins}\par}
\end{table}

\textcolor{black}{As discussed in the previous sections, our primary goal
is to determine the nonperturbative function of the CSS resummation formalism.
Hence, we need to include those experimental data, for which the nonperturbative
part dominates the transverse momentum distributions. This requirement suggests
using the low-energy fixed-target or collider Drell-Yan data in the region
where the transverse momentum $ Q_{T} $ of the lepton pair is much smaller
than its invariant mass $ Q $. Because the CSS formalism better describes
production of Drell-Yan pairs in the central rapidity region (as defined
in the center-of-mass frame of the initial-state hadrons), we shall concentrate
on the data with those properties. Based upon the above criteria, we chose
to consider data from the experiments listed in Table~\ref{data_table}
and in kinematical ranges shown in Table \ref{data_bins}. We have also examined
the E772 data \cite{E772} from the process $ p+H^{2}\ra \mu ^{+}\mu ^{-}+X $
at $ \sqrt{S}=56.6 $\,GeV and found them incompatible with the rest of
the data sets. Hence the E772 data were not included in the presented fits.}\footnote{%
For the \textcolor{black}{best-fit values of $ g_{i} $ given below, the
theory prediction for the E772 experiment is typically smaller than the data
by a factor of 2. Similarly, the E772 data are not well fit in the CTEQ global
analysis of parton distribution functions} \cite{wkt}.
} 

\textcolor{black}{The theoretical cross sections were calculated with the
help of the resummation package} \textsc{\textcolor{black}{Legacy}}\textcolor{black}{,
which was also used in previous fitting \cite{LY,BLLY} and
analytical studies \cite{oneW, wres, twoW1, twoW2, BalazsYuanHiggs, hw, qqhp},
as well as for generating input cross section
 grids for} \textsc{\textcolor{black}{ResBos}}
\textcolor{black}{Monte-Carlo integration program \cite{wres}. This package
is a high-performance tool for calculation of the resummed cross sections,
with the computational speed increased by up to a factor 800 after the reorganization
and translation of the source code into C/C++ programming language in 1999-2001.
During the preparation of this paper, we confirmed the stability of the numerical
calculation of the resummed cross sections (\ref{WY}) by comparing the output
of several Fourier-Bessel transform routines based on different algorithms
(adaptive integration, Fast Fourier-Bessel transform \cite{AndersonFFBT},
and Wolfram Research} \textcolor{black}{\emph{Mathematica 4.1}} \textsc{NIntegrate}
\textcolor{black}{function). Specifically, the outputs of three routines
are in a very good agreement at all values of $ Q_{T} $. For instance,
the $ Z $ boson cross sections presented in this paper and Ref.~\cite{BLLY}
are calculated with the relative numerical error less than $ 0.5\% $ at
$ Q_{T}\lesssim 50 $ GeV and less than $ 1-2\% $ at $ Q_{T}\gtrsim 50 $
GeV. Note that the relative error of about $ 1\% $ is comparable with
the size of higher-order (NNLO) corrections, as well as numerical uncertainties
in the existing two-loop PDF sets. More details on the tests of accuracy
of the resummation package will be presented elsewhere \cite{BalazsNadolskyYuan2002}.}\footnote{%
\textcolor{black}{An interface to the simplified version of} \textsc{\textcolor{black}{Legacy}}
\textcolor{black}{and online plotter of resummed transverse momentum distributions
are available on the Internet at http://hep.pa.msu.edu/wwwlegacy/ .}
} 

Using the above sets of the experimental data, we fit the
values of the nonperturbative parameters $ g_{1} $, $ g_{2} $ and $ g_{3} $
in the DWS-G form (\ref{DWS_form}), LY-G form (\ref{LY_form}), and BLNY
form (\ref{BLNY_form}) of the nonperturbative
 function $ \widetilde{W}_{j\bar{k}}^{NP}(b,Q,Q_{0},x_{1},x_{2}) $.
Since we allow the normalizations for the data
to float within the overall systematic normalization errors published by
the experiments, the best-fit values of $ g_{1} $, $ g_{2} $ and $ g_{3} $
are correlated with the best-fit values of the data normalization factors
$N_{fit}$ (individually applied to each data set). 
Note that the normalization of the CDF-$ Z $
Run 0 data was fixed to unity due to their poor statistics as compared to
the Run-1 data.

\begin{table}

\caption{The results of the fits. Here, \protect$ N_{fit}\protect$
is the fitted normalization for each experiment.\footnote{
Thus, by definition, NORM in  Ref.~\cite{BLLY} is equal to $1/N_{fit} \protect$. 
}}

{\centering \begin{tabular}{|c|c|c|c|}
\hline
Parameter &
DWS-G fit&
 LY-G fit&
 BLNY fit \\
\hline
\hline
$ g_{1} $&
 0.016 &
 0.02 &
 0.21 \\
\hline
$ g_{2} $&
 0.54 &
 0.55 &
 0.68 \\
\hline
$ g_{3} $&
 0.00 &
-1.50 &
-0.60 \\
\hline
\hline
CDF $ Z $ Run-0 &
 1.00 &
 1.00 &
 1.00 \\
 $ N_{fit}$ &
\textcolor{black}{(fixed)}&
\textcolor{black}{(fixed)}&
\textcolor{black}{(fixed)}\\
\hline
\hline
R209 &
 1.02 &
 1.01 &
 0.86 \\
 $ N_{fit}$ &
&
&
\\
\hline
E605 &
 1.15 &
 1.07 &
 1.00 \\
 $ N_{fit}$ &
&
&
\\
\hline
E288 &
 1.23 &
 1.28 &
 1.19 \\
 $ N_{fit}$ &
&
&
\\
\hline
\D0 $ Z $ Run-1 &
 1.01 &
 1.01 &
 1.00 \\
 $ N_{fit}$ &
&
&
\\
\hline
CDF $ Z $ Run-1 &
 0.89 &
 0.90 &
 0.89 \\
 $ N_{fit}$ &
&
&
\\
\hline
\hline
$ \chi ^{2} $&
 416 &
 407 &
 176 \\
\hline
$ \chi ^{2}/{\textrm{dof}} $&
 3.47 &
 3.42 &
 1.48  \\
\hline
\end{tabular}\par}

{\centering \label{fit_result}\par}
\end{table}

\begin{figure}
{\centering \resizebox*{12cm}{4in}{\includegraphics{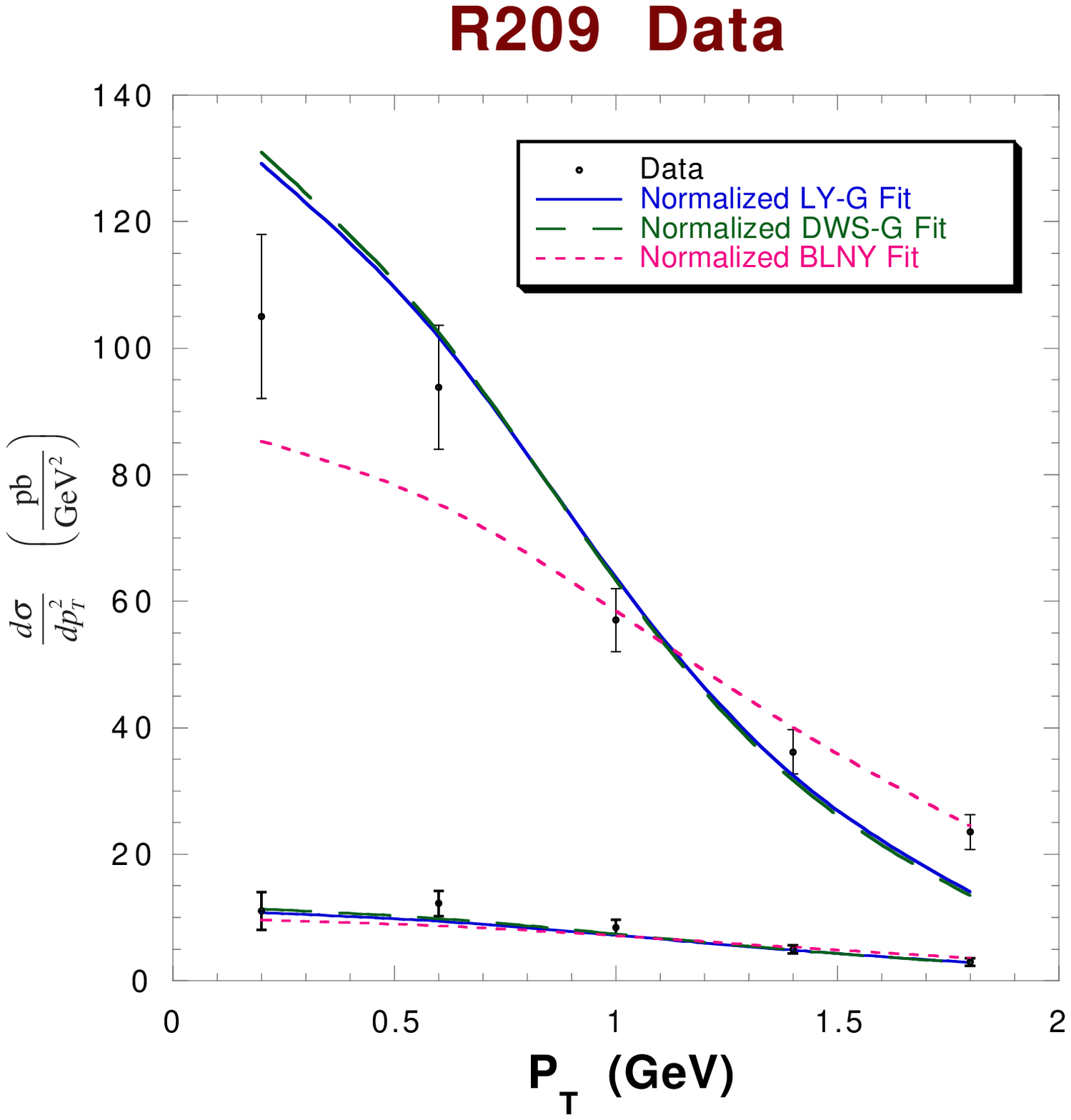}} \par}

\caption{\label{fig:r209} Comparison to the R209 data for the process
\protect$ p+p\ra \mu ^{+}\mu ^{-}+X\protect $
process at \protect$ \protect \sqrt{S}=62\protect $\,GeV. The data are
the published experimental values. The curves are
the results of the fits and are multiplied by the best-fit values of 
$1/N_{fit}$ given in Table III.}
\end{figure}

\begin{figure}
{\centering \resizebox*{12cm}{4in}{\includegraphics{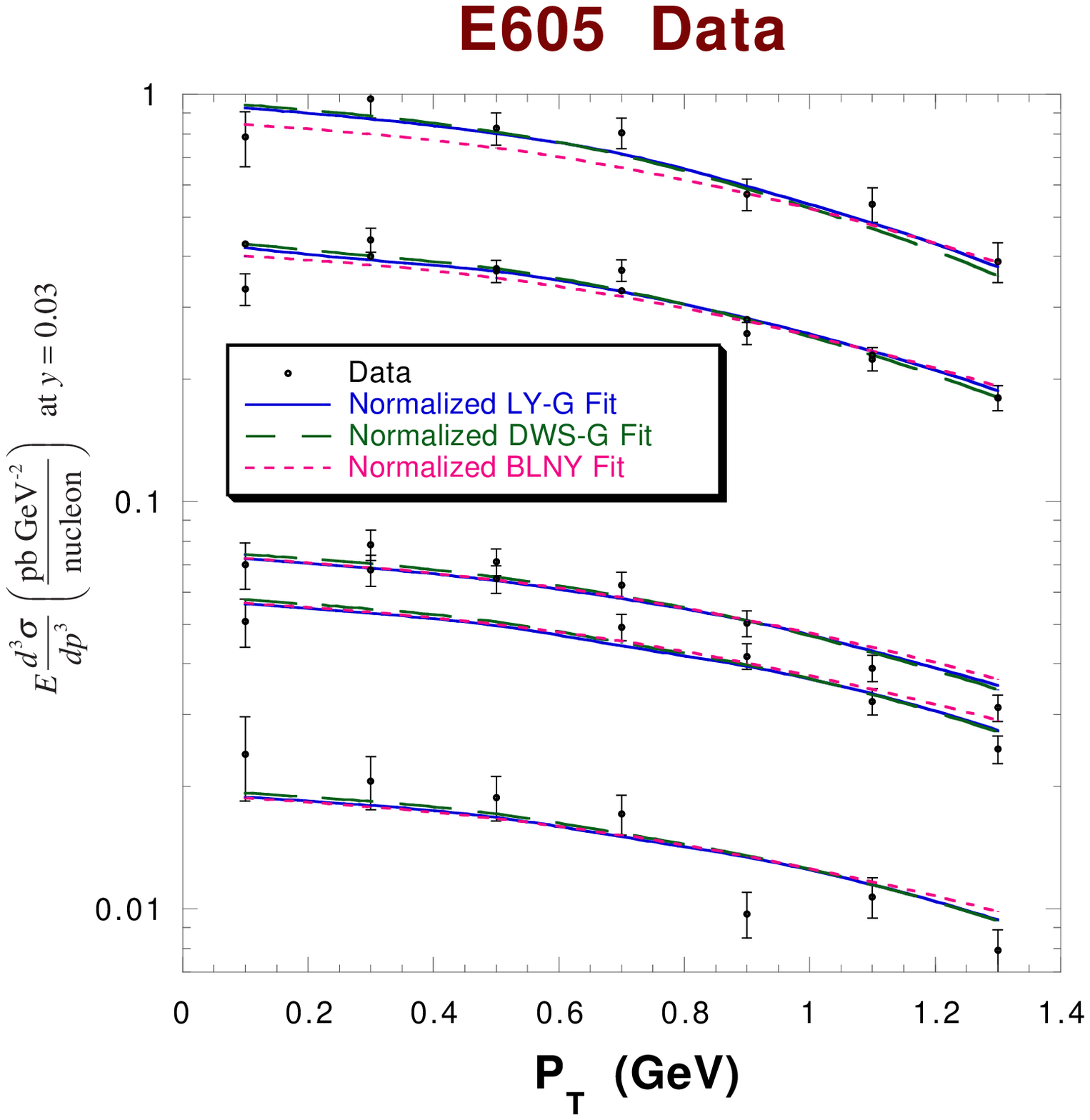}}  \par}

\caption{\label{fig:e605} Comparison to the E605 data for the
process \protect$ p+Cu\ra \mu ^{+}\mu ^{-}+X\protect $
at \protect$ \protect \sqrt{S}=38.8\protect $\,GeV. The data are
the published experimental values. The curves are the
results of the fits multiplied by the best-fit values of $1/N_{fit}$
given in Table III.}
\end{figure}

\begin{figure}
{\centering \resizebox*{12cm}{4in}{\includegraphics{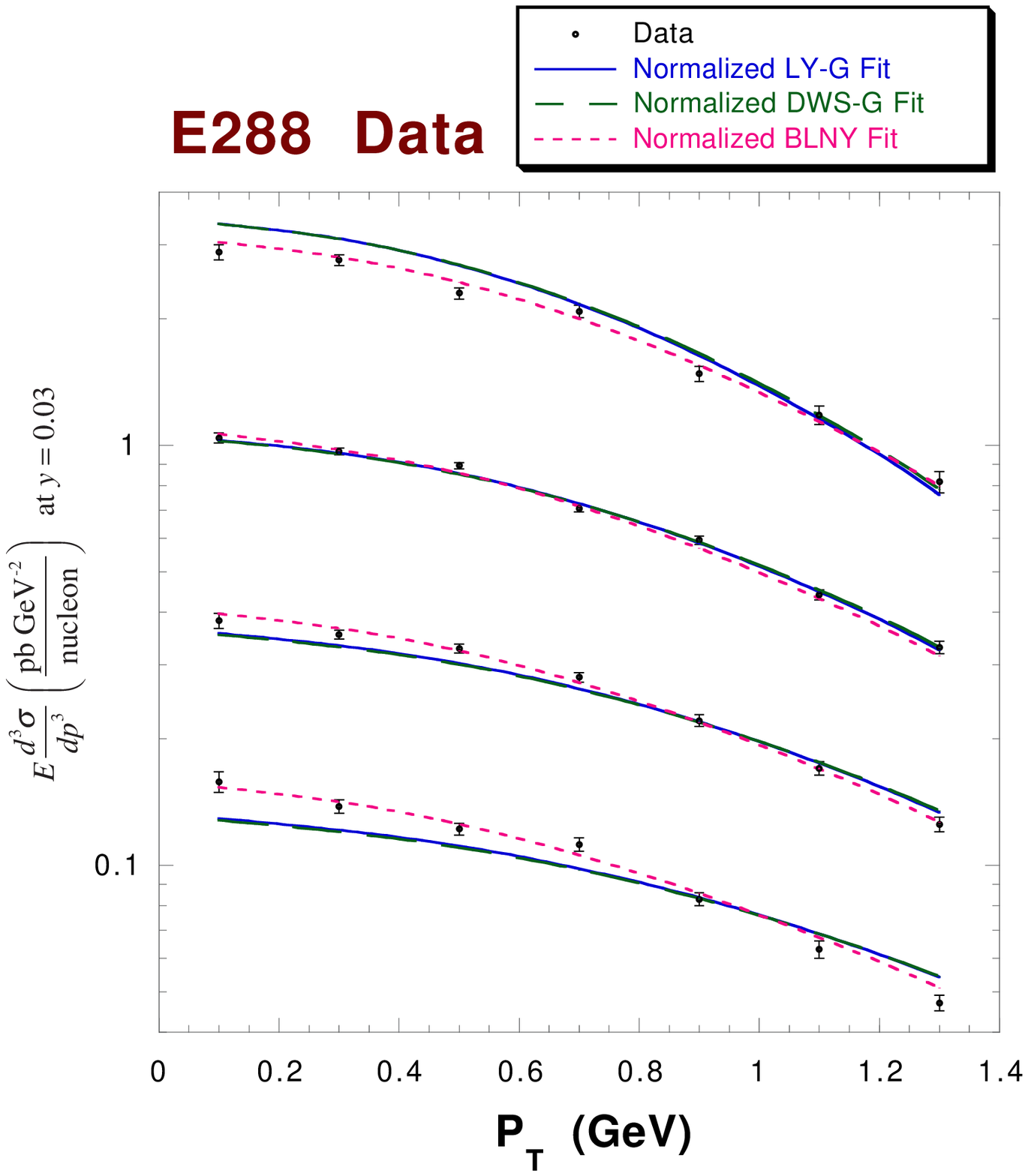}}  \par}

\caption{ Comparison to the E288 data for the process
\protect$ p+Cu\ra \mu ^{+}\mu ^{-}+X\protect $
at \protect$ \protect \sqrt{S}=27.4\protect $\,GeV. The data are
the published experimental values. The curves are
the results of the fits and are multiplied by the best-fit values of 
$1/N_{fit}$ given in Table III.
\label{fig:e288}}

\end{figure}

\begin{figure}
{\centering \resizebox*{12cm}{4in}{\includegraphics{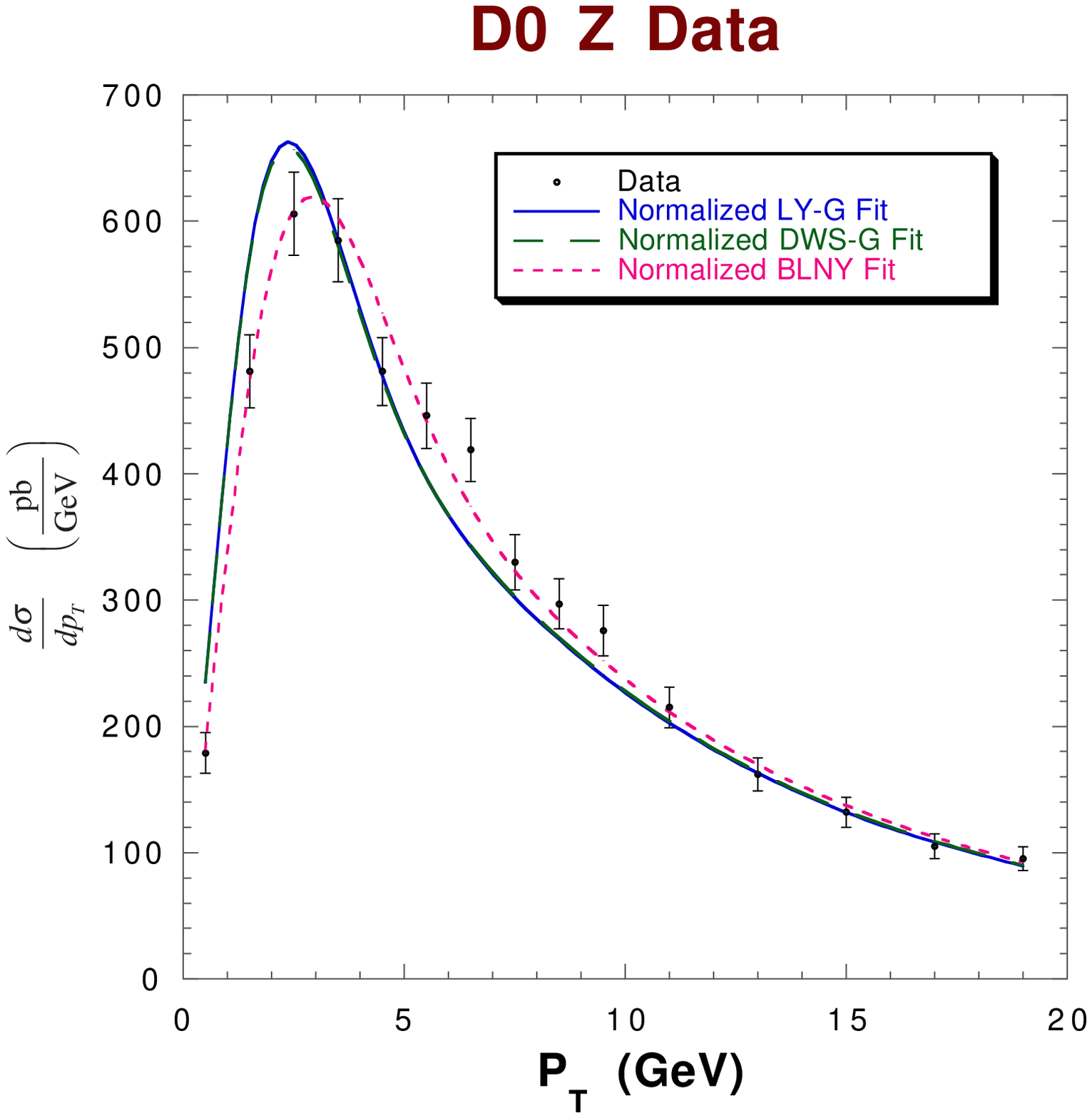}}  \par}

\caption{\label{fig:d0Z} Comparison to the \D0-\protect$ Z\protect $ Run-1 data. The data are the published experimental values.
The curves are 
the results of the fits and are multiplied by the best-fit 
values of $1/N_{fit}$ given in Table III.}
\end{figure}

\begin{figure}
{\centering \resizebox*{12cm}{4in}{\includegraphics{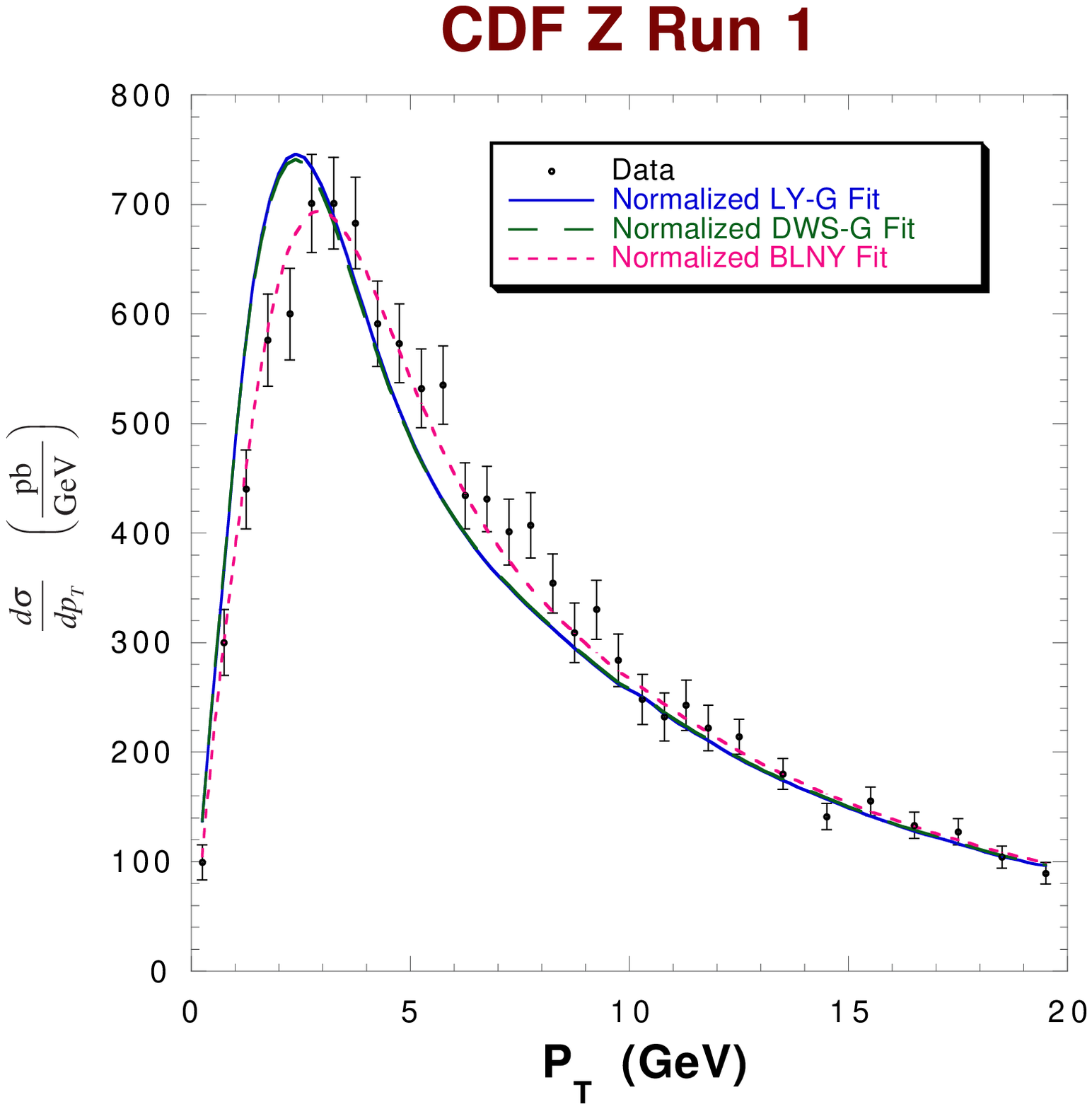}}  \par}

\caption{Comparison to the CDF-\protect$ Z\protect $ Run-1 data. 
The data are the published experimental values. The curves are
the results of the fits and are multiplied by the best-fit value of $1/N_{fit}$ given in Table III.
\label{fig:cdfZ}}
\end{figure}

\textcolor{black}{Table~\ref{fit_result} summarizes our results. To illustrate
the quality of each fit, Figs.~\ref{fig:r209}-\ref{fig:cdfZ} compare theory
calculations for the DWS-G, LY-G, and BLNY parameterizations to each data
set. We emphasize again that the new LY-G parameterization presented in Table~\ref{fit_result}
was obtained by applying the conventional global fitting procedure to the
enlarged data set listed in Tables~\ref{data_table} and \ref{data_bins}.
In contrast, the original LY fit in Ref.~\cite{LY} was obtained by first
fitting the $ g_{2} $ parameter using the CDF-$ Z $ Run 0 and R209
data, and then fitting $ g_{1} $ and $ g_{3} $ parameters after including
the other available Drell-Yan data (which is a small subset of the data in
the current fit).}

\textcolor{black}{It is evident that the Gaussian BLNY parameterization fits
the whole data sample noticeably better than the other two parameterizations,
both in terms of $ \chi ^{2} $ per degree of freedom (dof) given in Table~\ref{fit_result}
and in terms of the pictorial comparison in Figs.~\ref{fig:r209}-\ref{fig:cdfZ}.
When compared to the Run-1 $ Z $ data, the DWS-G and LY-G fits both fail
to match the height and position of the peak in the transverse momentum distribution
(Figs.~\ref{fig:d0Z} and \ref{fig:cdfZ}). Similarly, according to Fig.~\ref{fig:e288},
the BLNY parameterization leads to the best agreement with the E288 data.
The three fits are indistinguishable when compared to the E605 data, cf.
Fig.~\ref{fig:e605}. Fig.~\ref{fig:r209} shows a clear difference between
the BLNY fit and the other two fits for the lowest mass bin of the R209 data
(the upper data in the Figure). However, the $ \chi ^{2} $ contribution
from this mass bin is about the same for all three fits.}
\begin{figure}
{\centering \resizebox*{!}{9cm}{\includegraphics{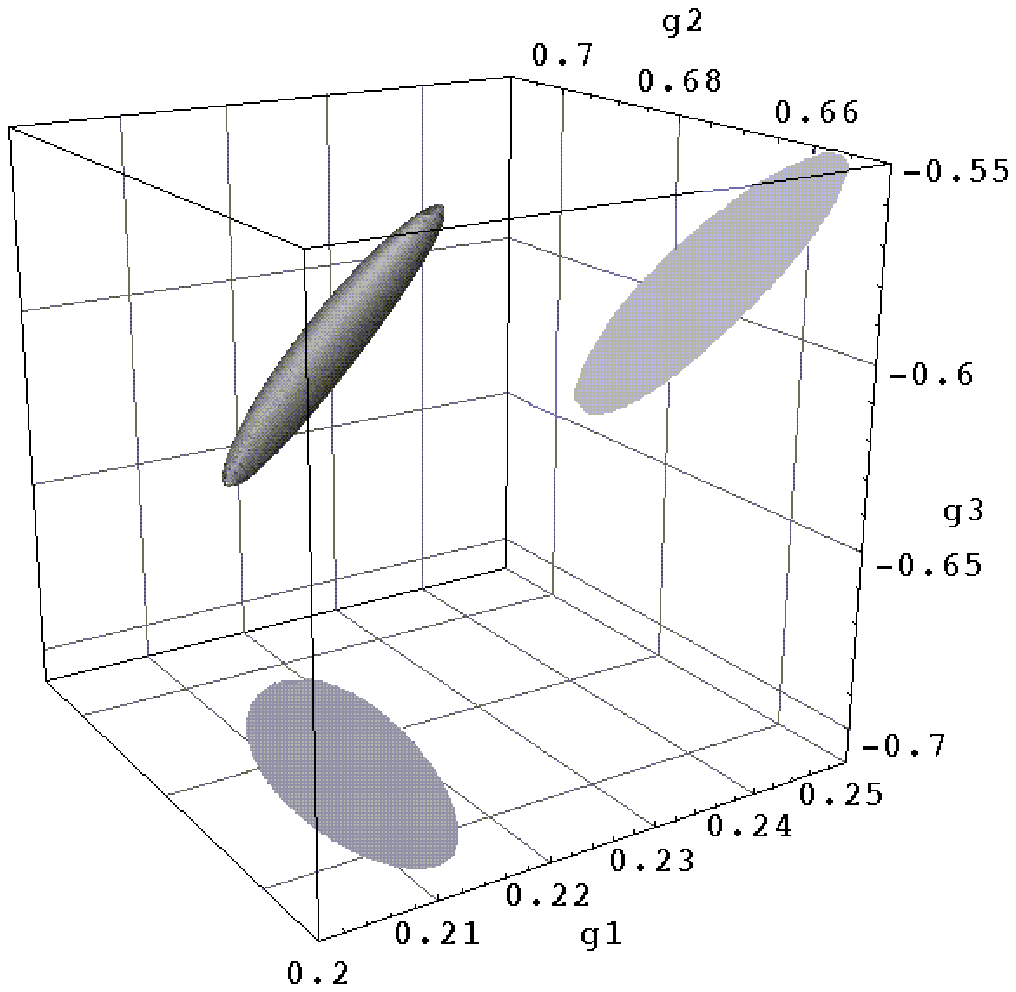}} \par}

\caption{Uncertainty contour and two dimensional projections for the 3-parameter
Gaussian BLNY form fit, cf. Eq.~(\ref{BLNY_form}). }

\label{fig:3dg}
\end{figure}

\textcolor{black}{The error on the fitted nonperturbative parameters $ g_{1} $,
$ g_{2} $ and $ g_{3} $ can be calculated by examining the $ \chi ^{2} $
distribution of the fit. For the BLNY form \begin{equation}
\label{BLNY_form_2}
\widetilde{W}_{j\bar{k}}^{NP}(b,Q,Q_{0},x_{1},x_{2})={\textrm{exp}}
\left[ -g_{1}(1+g_{3}\ln {(100x_{1}x_{2})})-g_{2}
\ln \left( {\frac{Q}{2Q_{0}}}\right) \right] b^{2}
\end{equation}
 with $ Q_{0}=1.6\mbox {\, GeV} $ and $ b_{max}=0.5\mbox {\, GeV}^{-1} $,
we found that \begin{equation}
\label{BLNY_g}
g_{1}=0.21_{-0.01}^{+0.01}\mbox {\, GeV}^{2}\; ,\; g_{2}=
0.68_{-0.02}^{+0.01}\mbox {\, GeV}^{2}\; ,\; g_{3}=-0.6_{-0.04}^{+0.05}.
\end{equation}
The errors in Eq.~(\ref{BLNY_g}) were computed as follows. First, $ \chi ^{2} $
values were calculated around their minimum $ \chi ^{2}_{min} $ in order
to obtain, in essence, a three dimensional function
$ \chi ^{2}(g_{1},g_{2},g_{3}) $.}\footnote{%
In this calculation, we scanned the values of $ g_{1} $ and $ g_{2} $
between 0 and 1, and $ g_{3} $ between $ -2 $ and 3.
} \textcolor{black}{Next, we plotted an ellipsoid surface determined by the
condition $ \chi ^{2}(g_{1},g_{2},g_{3})=\chi ^{2}_{min}+1 $ in the three-dimensional
space of parameters $ g_{i} $, cf.~Fig.~\ref{fig:3dg}. The extremes
in each coordinate for this surface were taken as the errors for the respective
parameters. Finally, we note that using $ \chi ^{2}_{min}+1 $ as the confidence
limit for determining the values of $ g_{i} $ in the presence of substantial
systematic errors is generally idealistic, for the experiments often make
judgments on systematic uncertainties that are not of a Gaussian nature.
Further discussion of this issue can be found in the next Section.}

\begin{figure}
{\centering \resizebox*{!}{7cm}{\includegraphics{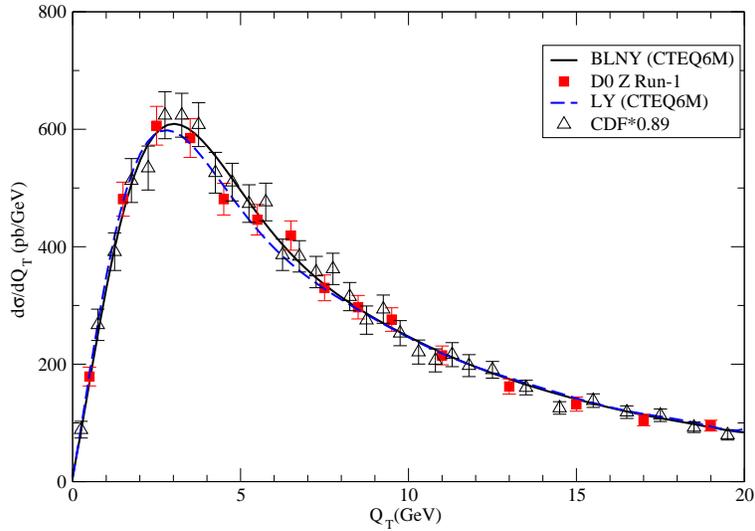}} \par}

\caption{Transverse momentum distributions of \protect$ Z\protect $
bosons at 
the Tevatron Run-1. The theory curves, calculated using the CTEQ6M
PDF's \cite{cteq6} and the BLNY parameterization (solid line)
or the original LY parameterization, shown as dashed line, are
compared to the \D0 data and CDF data. The data curves have
been multiplied by an overall normalization factor 1.0 in \D0
data and 0.9 in CDF data.}

\label{fig:qtz_cteq6}
\end{figure}

\section{Discussion}

\textcolor{black}{To fit the complete set of the experimental data (more
than 100 data points), this analysis introduced 3 free parameters ($ g_{1} $,
$ g_{2} $ and $ g_{3} $) in the parameterization of $ \widetilde{W}_{j\bar{k}}^{NP} $,
together with the chosen values of the parameters $ Q_{0} $ and $ b_{max} $.
In this paper and Ref.~\cite{BLLY}, we have chosen $ Q_{0}=1.6 $ GeV,
which coincides with the lowest energy scale used in the CTEQ3M PDF set.
This choice is nothing more than a matter of convenience, since it enforces
positivity of the logarithm $ \ln (Q/2Q_{0}) $ in the $ Q $ range of
the fitted data ($ Q\geq 5 $ GeV). But $ Q_{0} $ does not have to take
that special value, since its variations can be compensated for in the full
nonperturbative function $ \widetilde{W}_{j\bar{k}}^{NP} $ by adjusting
parameters $ g_{1} $ and $ g_{3} $ (at $ g_{2} $ fixed). For instance,
we could have chosen $ Q_{0} $ to be equal to $ 1/b_{max} $ and rewrite
Eq.~(\ref{BLNY_form}) as\begin{equation}
\label{BLNY_form3}
\widetilde{W}_{j\bar{k}}^{NP}(b,Q,Q_{0},x_{1},x_{2})={\textrm{exp}}
\left[ -g^{'}_{1}(1+g^{'}_{3}\ln {(100x_{1}x_{2})})-g^{'}_{2}\ln (Qb_{max})\right] b^{2},
\end{equation}
 where $ g^{'}_{2}=g_{2} $, $ g^{'}_{1}=g_{1}-g_{2}\ln (2Q_{0}b_{max}) $,
and $ g^{'}_{3}=g_{3}g_{1}/g^{'}_{1} $. This new form of $ \widetilde{W}^{NP}_{j\bar{k}} $
would be completely identical to the original form in Eq.~(\ref{BLNY_form}).
Hence, the total number of parameters needed to describe $ \widetilde{W}_{j{\bar{k}}}(b) $
in the used prescription is four,} \textcolor{black}{\emph{i.e.,}} \textcolor{black}{$ g_{i} $
and $ b_{max} $. }

\textcolor{black}{While the parameter $ Q_{0} $ plays no dynamical role,
the meaning of the parameter $ b_{max} $ is quite different. Roughly speaking,
its purpose is to separate nonperturbative effects from perturbative contributions
through the introduction of the variable $ b_{*} $ defined in Eq.~(\ref{bstar}).
According to its definition, the variable $ b_{*} $ is practically equal
to $ b $ when $ b^{2}\ll b^{2}_{max} $. For $ b\rightarrow \infty  $,
it asymptotically approaches $ b_{max} $. Hence, in the factorized CSS
representation (\ref{wpres}) the perturbative part $ \widetilde{W}^{pert}_{j{\bar{k}}}(b_{*}) $
approaches its exact value (evaluated at $ b $) when $ b\rightarrow 0 $,
and it is frozen at $ b_{*}\approx b_{max} $ when $ b\rightarrow \infty  $.
While $ b_{max} $ should lie in the perturbative region to make the computation
of $ \widetilde{W}^{pert}_{j\bar{k}}(b_{*}) $ feasible, it is also desirable
to make it as large as possible, to reduce deviations of $ \widetilde{W}^{pert}_{j\bar{k}} $
from its exact behavior at smaller $ b $. Note, however, that the changes
in the behavior of $ \widetilde{W}^{pert}_{j\bar{k}} $ due to the $ b_{*} $
prescription, such as the loss of suppression of $ \widetilde{W}^{pert}_{j\bar{k}} $
by the perturbative Sudakov factor at $ b\gtrsim b_{max} $, can be compensated
for by increasing the magnitude of the nonperturbative function $ \widetilde{W}^{NP}_{j\bar{k}} $.
Hence, in the $ b_{*} $ prescription the nonperturbative function $ \widetilde{W}^{NP}_{j\bar{k}} $
generally serves a dual purpose of the parameterization for truly nonperturbative
effects and compensation factor for modifications in $ \widetilde{W}^{pert}_{j\bar{k}} $
due to the $ b_{*} $ variable. Consequently, the best-fit parameterization
of $ \widetilde{W}^{NP}_{j\bar{k}} $ depends on the choice of $ b_{max} $. }

%The above requirements for $ b_{max} $ suggest to choose
%it of order $ 1/(1\, {\rm GeV}), $ such as $ b_{max}=0.5\mbox {\, GeV}^{-1} $
%used in our fits. By utilizing this choice, we found that the Tevatron $ Z $
%data plays an important role in the determination of $ \widetilde{W}^{NP}_{j\bar{k}}, $
%in accordance with our expectations from earlier studies. Recently, the correctness
%of this conclusion, as well as importance of global fits for the large-$ Q $
%physics were questioned by recent papers \cite{qiu1}, which argued that
%heavy boson production is not sensitive to the nonperturbative input. These
%papers suggested that the sensitivity of the large-$ Q $ cross sections
%to $ \widetilde{W}^{NP}_{j\bar{k}} $ is an artifact of the $ b_{*} $
%approach. As an alternative, 

\textcolor{black}{
Based on the fact that the $ b_{*} $ prescription with the BLNY form of
the non-perturbative function provides an excellent fit 
to the whole set of Drell-Yan-like data,\footnote{
For example, as shown in Figs.~\ref{fig:d0Z}
and \ref{fig:cdfZ}, both the peak position and the shape of the 
transverse momentum
distribution of the $Z$ boson measured by the \D0 and CDF Collaborations
at the Run-1 of the Tevatron are very well described by the theory
calculations. This feature is extremely important for the precision 
measurement of the $W$ boson mass at the Tevatron.
}
 we conclude that the $ b_{*} $
prescription remains an adequate method for studies
of resummed $ Q_{T} $ distributions.}%\footnote{%
%Discussion of details of the $ b_{*} $ method, such as the role of
%the nonperturbative contributions, high-$ Q_{T} $
%behavior, and energy dependence, will be included in another upcoming paper
%\cite{BalazsNadolskyYuan2002}.
%} 
\textcolor{black}{This, however, does not mean that future tests and improvements
of this method will not be necessary.}
For instance, one might consider a larger value of $b_{max}$ for the
fit. Up to now, all the fits performed in the literature have taken 
$b_{max}$ to be $0.5   {\, \mbox{GeV}}^{-1}$.
However, $b_{max}$ could be chosen to be as large as 
$ b_{0}/\mu _{F}^{0} $, where $ b_{0}=2e^{-\gamma _{E}}\approx 1.122... $,
and $ \mu _{F}^{0} $ is the initial energy scale for the PDF set in use.
Several of the most recent PDF sets, such as CTEQ6M \cite{cteq6}, MRST'2001
\cite{MRST2001} and MRST'2002 \cite{MRST2002}, use $ \mu _{F}^{0} $
as low as 1~GeV. Correspondingly, for the new PDF sets $ b_{max} $ can
be as large as $ 1.122\mbox {\, GeV}^{-1} $.
It will be interesting to test the $ b_{*} $ prescription
with such increased value of $ b_{max} $ in a future study.
Moreover, an approach alternative to the $b_*$ prescription was 
recently proposed in Refs.~\cite{qiu1,qiu2}.
Fig.~7 of Ref.~\cite{qiu2} shows the comparison   
of this new theory calculation to the Run-1 CDF data, which should be
compared to Fig.~\ref{fig:cdfZ} of this paper.
Furthermore, in Refs.~\cite{joint,joint2}, another method for
performing the extrapolation to the large-$b$ region was proposed.
To see the difference between the theory predictions 
for the Tevatron $Z$ data, one can compare Figs.~5 and 6 of Ref.~\cite{joint2}
to Figs.~5 and 4  of this paper. In a forthcoming publication 
\cite{BalazsNadolskyYuan2002}, 
we shall present a detailed comparison of various methods for description 
of large-$b$ physics to the Tevatron $Z$ data.

\textcolor{black}{We would like to conclude this section with a remark about
the dependence of our results on the choice of parton distribution functions.
As noted above, the fitted nonperturbative function $ \widetilde{W}_{j{\bar{k}}}^{NP}(b) $
in the CSS resummation formalism is correlated to the PDF's used in the theory
calculation. In the current fit, we have chosen to use CTEQ3M PDF's to facilitate
the comparison with the previous results in Ref.~\cite{BLLY}. The usage
of a more modern PDF set with the BLNY parameterization of $ \widetilde{W}^{NP}_{j\bar{k}} $
would result in a difference of a few percent in $ Q_{T} $ distributions.
Currently, such differences are of order of normalization errors quoted by
the experiments. Hence, good agreement between the theory and data can be
obtained by small normalization shifts in the data. We illustrate this point
in Fig.~\ref{fig:qtz_cteq6}, which compares the Tevatron Run-1 $ Z $
data to the CSS resummation calculation performed using CTEQ6M PDF's \cite{cteq6}
and BLNY parameterization (\ref{BLNY_form}) with the best-fit nonperturbative
parameters (\ref{BLNY_g}). By adjusting the normalization of the CDF data by
the best-best value $N_{fit}=0.89$ in Fig.~\ref{fig:qtz_cteq6},
the theory is brought in a good agreement with both sets of the
Run-1 $Z$ data. }
For comparison, we also show in Fig.~\ref{fig:qtz_cteq6}
the prediction from using CTEQ6M PDF's and
the original LY parametrization \cite{LY}.

\textcolor{black}{As the quality of the data improves in the future, the
correlation between the nonpertubative function $ \widetilde{W}_{j{\bar{k}}}^{NP}(b) $
and PDF's will become more important. Hence, it will be certainly desirable
to repeat the work done in this paper using the most recent set of PDF's
and potentially perform the joint global analysis of the PDF's and CSS nonperturbative
function. Furthermore, as the statistical errors decrease, correct treatment
of the systematic errors (which are hardly of a Gaussian nature in many experiments)
becomes ever more necessary. Therefore, a method more elaborate than the
simple criterion $ \chi ^{2}_{min}+1 $ used in the current analysis will
be needed to determine the true confidence limit for the nonperturbative
function. Recently, new efficient methods were proposed to perform the error
analysis on the nonperturbative parameters in the PDF's in the presence of
systematic errors \cite{pdferr, cteq6, MRST2002}. In the future, the same
methods can also be applied to determine the errors for the nonperturbative
parameters $ g_{1} $, $ g_{2} $, and $ g_{3} $ in
the BLNY nonperturbative function.}

\section{Conclusions}

\textcolor{black}{We have shown for the first time that the complete set
of low energy Drell-Yan data (R209, E605 and E288) and the Tevatron Run-1
$ Z $-boson data can be}
 \textit{\textcolor{black}{simultaneously}} \textcolor{black}{described
by the CSS resummation formalism. 
This is the first truly global fit, which treats 
all the low energy Drell-Yan data and the Tevatron Z data on the same
footing.
This fact strongly supports the universality
of the CSS nonperturbative function $ \widetilde{W}_{j\bar{k}}^{NP} $.
Just as the universality of the PDF's allows us to predict the inclusive
rate for a scattering process in hadron collisions, the
universality of $ \widetilde{W}_{j\bar{k}}^{NP} $
allows us to predict the transverse momentum distributions in production
of single vector bosons and other similar processes at hadron colliders.
For example, if true, the resummation formalism can be applied not only to
the Drell-Yan pair production and $ W $ (or $ Z $) boson production,
but also to associated production of Higgs and $ W $ (or Higgs and $ Z $)
bosons \cite{hw}, diphoton production \cite{twoW1}, $ ZZ $ (or $ WW $)
pair production \cite{twoW2}, and $ s $-channel neutral or charged Higgs
boson production (induced from quark fusion with large Yukawa coupling to
Higgs boson) \cite{qqhp}. When these processes are measured to a good accuracy,
they can be used to further test the CSS resummation formalism. Just as the
PDF's must be constantly refined in order to fit new experimental data, the
nonperturbative function $ \widetilde{W}_{j\bar{k}}^{NP} $ may also require
modification in the future. At the current stage, the existing data shows
remarkable preference for a Gaussian form of the nonperturbative function
with $ Q $ dependence predicted by the CSS formalism.}

\section*{\textcolor{black}{Acknowledgments}}

\textcolor{black}{We thank C.~Bal\'{a}zs, F. Olness, and the CTEQ collaboration
for useful discussions. The presented fits were included in F.~Landry's
Ph.~D.~thesis \cite{LandryThesis}. The work of R.~B. and C.-P.~Y. was
supported in part by National Science Foundation grants 
PHY-0140106 and PHY-PHY-0100677,
respectively. The work of P.~M.~N. was supported by the U.S. Department
of Energy and Lightner-Sams Foundation.} 

\end{document}